\documentclass[10pts]{article}
\usepackage{amsthm}
\theoremstyle{plain}
\usepackage{graphicx} 
\usepackage{braket}
\usepackage{amsmath}
\usepackage{epstopdf}
\usepackage{cancel}
\usepackage{cite}

\newtheorem*{theorem*}{Theorem}
\usepackage{amsfonts}
\begin{document}

\title{\bf A Relativistic One-Particle Time of Arrival Operator for a Free Spin-$1/2$ Particle in $(1+1)$ Dimensions}
\author{Joseph Bunao\thanks{Corresponding Author: jbunao@nip.upd.edu.ph} and Eric A. Galapon\thanks{eagalapon@up.edu.ph, eric.galapon@upd.edu.ph}  \\ Theoretical Physics Group, National Institute of Physics\\University of the Philippines, 1101 Philippines}
\maketitle
\begin{abstract}
As a follow-up to a recent study in the spin-$0$ case [J. Bunao and E. A. Galapon, Ann. Phys. \textbf{353}, 83-106 (2015)], we construct a one-particle Time of Arrival (TOA) operator conjugate to a Hamiltonian describing a free relativistic spin-$1/2$ particle in one spatial dimension. Upon transformation in a representation where the Hamiltonian is diagonal, it turns out that the constructed operator consists of an operator term $\mathcal{\hat{T}}$ whose action is the same as in the spin-$0$ case, and another operator term $\mathcal{\hat{T}}_{0}$ which commutes with the Hamiltonian but breaks invariance under parity inversion. If we must impose this symmetry on our TOA operator, then we can throw away $\mathcal{\hat{T}}_{0}$ so that the TOA operator is just $\mathcal{\hat{T}}$.
\end{abstract}

\section{Introduction}
\label{intro}
The marriage of relativity and quantum mechanics is a difficult one. Even for the case of special relativity, the interpretations can be somewhat troublesome. In the regime where the effects of quantum mechanics and special relativity are no longer negligible, the particle number in a system can, and will, fluctuate. Roughly speaking, this is because mass is just another form of energy, according to special relativity, and energy can fluctuate, according to quantum mechanics. Thus, there can be massive particles being created and annihilated in a relativistic quantum system. This poses a problem on the interpretation of the supposedly one-particle wavefunctions satisfying relativistic wave equations (i.e. the Klein-Gordon and Dirac Equations) and the quantum operators representing the physical observables of the system. To illustrate, consider a time of arrival (TOA) experiment. Suppose we start with a particle in some initial state and let it propagate towards a detector. When a particle arrives at the detector, it records the time of arrival. However, how can we be sure that the particle that arrived is the same one we started with? This is not an issue in the usual quantum mechanics but it is a question that must be addressed in the relativistic quantum regime. We can also repeat the same experiment with particles starting with the same initial states in order to get a TOA probability distribution. How can we then interpret this distribution not knowing whether the particles that have arrived may not be the ones we have started with? Theoretically, this also translates into a problem for the quantum operators in which probability distributions are to be derived from. 

In \cite{art, book}, the one-particle interpretation of relativistic quantum mechanics was reinforced by interpreting the wavefunction of a single free particle to be one of the two independent ('positive' or 'negative') solutions of the Klein-Gordon equation (for spin-0 particles) or the Dirac equation (for spin-1/2 particles). One-particle analysis would then involve restricting ourselves into a Hilbert (sub)space spanned by the positive or negative states and working with true one-particle quantum operators which do not map these two subspaces into each other. They have calculated several one-particle operators for, say, energy (the Hamiltonian), momentum, and position. However, a time operator was not constructed implying the parametric role of time in the one-particle formalism. This leaves us a certain sense of disparity with regards to the role of time in relativistic quantum mechanics. That is, in special relativity, time is a quantity that is intrinsic to the system in question. Moreover, space and time are intertwined so that they are treated on the same footing. However, in the earlier days of quantum mechanics, time is not treated as a physical observable quantity but only an external and absolute parameter marking the evolution of a system. This is in contrast with the observable status of space. The powerful formalism of quantum field theory (QFT), which states that particles and antiparticles are just excitations of a quantum field subject to fluctuations, demotes the role of space so that time and space are on equal footing. They both, however, are just parameters labeling the quantum field operators. 

Going in a different direction, there are studies in both non-relativistic and relativistic quantum mechanics promoting the role of time to an observable 
\cite{grot, abohm, gal1, gal2, gal3, gal4, gal5, gal6, gal7, muga1, muga2, muga3, prot, gpd, disct, atrqm, rfmtoa}. Particularly, it was calculated that the TOA operator for a non-relativistic particle with mass $m_{0}$ is $-m_{0}\hat{T}_{-1,1}$ \cite{abohm, gal1}, where the $\hat{T}_{m,n}$'s form a complete and linearly independent set called the Bender-Dunne operators \cite{ben1, ben2}. In \cite{prot}, a proper time operator was found for a relativistic electron and was used to study some properties of several position operators. And in \cite{gpd}, a method for constructing TOA probabilities valid for any experimental setup (including relativistic systems with interactions described by QFT) was developed. They associated the TOA of a particle as the time instant when there is a transition in the degrees of freedom of the detector. While in \cite{disct}, a relativistic particle moving in (5+1) dimensions, where two of the spatial dimensions are compactified into a torus, was studied. They constructed a discrete physical time based on a quasi-local invariant observable. It was then found to be related to the proper time on average. Another study \cite{atrqm, rfmtoa} is on the construction and properties of a self-adjoint relativistic TOA operator for a free spin-$1/2$ particle. They found that the self-adjointness was due to the existence of particle and antiparticle solutions of the Dirac Equation. An important concept that has risen from these studies is the idea of \emph{supraquantization} \cite{gal3}. It is the idea of constructing quantum observables without quantization. This is so that we can break from the circularity of promoting a classical quantity into a quantum operator then checking if it indeed reduces to the said quantity in the classical limit. That is, if classical mechanics can be thought of as only a limiting case of quantum mechanics, then quantum mechanics must be autonomous from classical mechanics. Quantum operators should then be constructed from quantum mechanical principles with the added condition that the corresponding classical quantities indeed arise in the appropriate limit. TOA operators are then constructed as being canonically conjugate with the system Hamiltonian. For a free particle, the TOA operator is indeed $-m_{0}\hat{T}_{-1,1}$ \cite{gal1, gal2}. For potentials of higher order than quadratic in position, the usual quantization of the classical TOA does not produce an operator canonically conjugate with the Hamiltonian. Such instances are the shortcomings of quantization. However, we can construct a TOA operator that is conjugate with the Hamiltonian (from a quantum principle) and see that it would reduce to the correct classical TOA \cite{gal1, gal2}. Likewise, we should then be able to construct a TOA operator for a relativistic particle that is conjugate with the system Hamiltonian without referring to any quantization rule. As stated earlier, this operator should be made into a true one-particle operator since we are in the relativistic quantum regime. 

This study is then a follow-up to \cite{kgrqtoa}. That is, we will construct a relativistic one-particle TOA operator for a free spin-$1/2$ particle in one spatial dimension. Here, we will use similar methods done in \cite{kgrqtoa, art, book}, but for a free spin-$1/2$ particle. In Section \ref{bgrdd}, we briefly review the one particle formalism of the Dirac Equation and define the system Hamiltonian in the process. In Section \ref{constr}, we construct a TOA operator using a method similar to \cite{kgrqtoa} and put the resulting operator into a representation where the Hamiltonian is diagonal (the so-called $\Phi$-representation). In Section \ref{symm}, we impose that the TOA operator be invariant under parity inversion. And lastly in Section \ref{conc}, we conclude.

\section{One-Particle Interpretation of the Dirac Equation}
\label{bgrdd}

In $(1+1)$ dimensions, the free Dirac Equation may be written as
\begin{equation}\label{de}
i\hbar\frac{\partial \Psi}{\partial t} = -i\hbar c A \frac{\partial \Psi}{\partial x} + m_{0}c^{2}B\Psi
\end{equation}
where $A$ and $B$ are $2 \times 2$ hermitian matrices satisfying $\{A,B\} =AB+BA=0$ and $A^{2} = B^{2} = \left(\begin{smallmatrix} 1 & 0\\ 0 &1 \end{smallmatrix}\right)$, and $\Psi = (\psi_{1} \; \psi_{2})^{T}$ is a two-component column matrix describing a free spin-$1/2$ relativistic particle with mass $m_{0}$. The corresponding probability density is positive definite and can be readily constructed as $\rho = \Psi^{\dagger}\Psi = (\psi_{1}^{*} \; \psi_{2}^{*})(\psi_{1} \; \psi_{2})^{T} = |\psi_{1}|^{2} + |\psi_{2}|^{2}$. We then may define the Hamiltonian of the system as 
\begin{equation}
\mathcal{\hat{H}}_{\Psi} = -i\hbar c A \frac{\partial}{\partial x} + m_{0}c^{2}B
\end{equation}
which we call as being in the Schrodinger representation ($\Psi$-representation) as in \cite{kgrqtoa, book}. Acting twice on $\Psi$, we have
\begin{equation}
-\hbar^{2}\frac{\partial^{2} \Psi}{\partial t^{2}} = \mathcal{\hat{H}}_{\Psi}^{2} \Psi = -\hbar^{2}c^{2}\frac{\partial^{2} \Psi}{\partial x^{2}} + m_{0}^{2}c^{4}\Psi
\end{equation}
which means that each component of $\Psi$ satisfies the Klein-Gordon equation separately.

We also see that the solutions of Eq (\ref{de}) are of the form $\Psi \sim \exp(ipx/\hbar - iEt/\hbar)$ which tells us that the eigenvalules of $\mathcal{\hat{H}}_{\Psi}$ are $E = \lambda E_{p} = \pm E_{p} = \pm \sqrt{p^{2}c^{2}+m_{0}^{2}c^{4}}$. The Dirac Equation can then be taken as having two types of solutions characterized by $\lambda$, similar with the Klein-Gordon case. The difference is that, given a momentum $p$, the particle indeed can have an energy $+E_{p}$ or $-E_{p}$ \cite{book}. The interpretation provided by Dirac is that all of the negative energy states are occupied whereas the positive energy states are not. The transition of a particle from the 'sea of negative states' to a positive state leaves a hole in this sea which we now interpret as an antiparticle. The Dirac sea however necessarily requires an infinite number of particles so that a one-particle interpretation may not stand a rigorous treatment. Still, we insist on the interpretation by considering one-particle operators \cite{art, book}. First, we see that the Hamiltonian $\mathcal{\hat{H}}_{\Psi} = c A p +m_{0}c^{2}B$ (in momentum representation) can be diagonalized. Since $A = \sum_{j=1}^{3}\alpha_{j}\sigma_{j}$ and $B = \sum_{j=1}^{3}\beta_{j}\sigma_{j}$, where
\begin{displaymath}
\sigma_{0} = 
\begin{pmatrix}
 1 & 0\\
 0 & 1 \\
\end{pmatrix},
\sigma_{1} = 
\begin{pmatrix}
 0 & 1\\
 1 & 0 \\
\end{pmatrix},
\sigma_{2} = 
\begin{pmatrix}
 0 & -i\\
 i & 0 \\
\end{pmatrix},
\sigma_{3} = 
\begin{pmatrix}
 1 & 0\\
 0 & -1 \\
\end{pmatrix}
\end{displaymath}
we can use
\begin{equation}\label{utrans}
U = U^{\dagger} = U^{-1} = \frac{1}{\sqrt{2E_{p}(E_{p}+\alpha_{3}pc+\beta_{3}m_{0}c^{2})}}(E_{p}\tau_{3}+\mathcal{\hat{H}}_{\Psi})
\end{equation}
to show that
\begin{align}\label{tevol}
i\hbar\frac{\partial }{\partial t}(U\Psi) &= U \mathcal{\hat{H}}_{\Psi}U^{-1}(U\Psi) \nonumber\\
i\hbar\frac{\partial }{\partial t}\Phi &= \mathcal{\hat{H}}_{\Phi} \Phi = \sigma_{3} E_{p} \Phi .\nonumber\\
\end{align}
We see then explicitly that the eigenvalues of the Hamiltonian are $\pm E_{p}$. 

It is interesting to note that the time evolution of states generated by the diagonalized Dirac Hamiltonian, Eq (\ref{tevol}), is of the same form as the time evolution generated by the diagonalized Klein-Gordon Hamiltonian \cite{kgrqtoa,art,book}. Now, if we consider restricting our wavefunctions to be of the form $\Phi_{+} = (\phi_{+} \; 0)^{T}$ or $\Phi_{-} = (0 \; \phi_{-})^{T}$ then $\Phi_{\pm}$ has energy $\pm E_{p}$. Moreover, Eq (\ref{tevol}) states that the time evolution of $\Phi_{+}$ is independent of $\Phi_{-}$ and vice-versa. That is, the states do not mix as time evolves. We may then consider $\Phi_{\pm}$ as a one-particle wavefunction. True one-particle operators then should not map a positive state $\Phi_{+}$ into a negative state $\Phi_{-}$ and vice-versa. So if an operator $\mathcal{\hat{A}}_{\Psi}$ in the $\Psi$-representation is a true one-particle operator, then $\mathcal{\hat{A}}_{\Phi} = U \mathcal{\hat{A}}_{\Psi} U^{-1}$ in the so-called Feshbach-Villars representation ($\Phi$-representation) should be diagonal \cite{art, book}. An example of a one-particle operator is $\mathcal{\hat{H}}_{\Phi} = \sigma_{3} E_{p}$. These types of operators do not mix the positive and negative states.

The Hilbert space of states $\mathcal{H}$ may then be seen as being split into two subspaces $\mathcal{H}_{+}$ and $\mathcal{H}_{-}$ so that $\mathcal{H} = \mathcal{H}_{+} \oplus \mathcal{H}_{-}$. To be more specific, since we may write the inner product in $\mathcal{H}$ as $<\Psi_{1} | \Psi_{2}>  = \int \Psi_{1}^{\dagger} \Psi_{2} dp = \int \Psi_{1}^{\dagger}U^{\dagger}U\Psi_{2} dp = \int \Phi_{1}^{\dagger}\Phi_{2} dp = <\Phi_{1} | \Phi_{2}>$, then
\begin{displaymath}
\mathcal{H}_{\lambda} = \left\{ \Phi_{\lambda} = 
\begin{pmatrix}
\Theta (\lambda)\\
\Theta (-\lambda)
\end{pmatrix}
\phi_{\lambda}
\left|
<\Phi_{\lambda}|\Phi_{\lambda}> = \int_{-\infty}^{\infty}\Phi_{\lambda}^{\dagger}\Phi_{\lambda} dp \right.  = \int_{-\infty}^{\infty}|\phi_{\lambda}|^{2} dp <\infty \right\}
\end{displaymath}
with $\lambda = \pm 1$. If we, say, take $\Phi_{+} = (1\;0)^{T} \phi_{+}$ as describing a particle, then analysis of that same particle means that we are restricting ourselves in the state subspace $\mathcal{H}_{+}$ and are working with operators which do not mix the states of $\mathcal{H}_{+}$ and $\mathcal{H}_{-}$

\section{Constructing $\mathcal{\hat{T}}_{\Psi}$}
\label{constr}

In this section we attempt to construct a Time of Arrival operator $\mathcal{\hat{T}}_{\Psi}$ canonically conjugate with the system Hamiltonian $\mathcal{\hat{H}}_{\Psi}$ so that 
\begin{equation}\label{ccr}
\left[\mathcal{\hat{H}}_{\Psi}, \mathcal{\hat{T}}_{\Psi}\right]\Psi = i\hbar \Psi
\end{equation}
for any two-element column matrix $\Psi$, and the Hamiltonian is given by $\mathcal{\hat{H}}_{\Psi} = c A p +m_{0}c^{2}B$.
We proceed by expressing $\mathcal{\hat{T}}_{\Psi}$ in terms of the complete linearly independent set of basis operators $\hat{T}_{m,n}$ (so-called Bender-Dunne operators) \cite{ben1, ben2}, 
\begin{align}
\hat{T}_{m,n} &=\frac{1}{2^n}\sum_{k=0}^{n} \frac{n!}{k!(n-k)!} \hat{q}^{k} \hat{p}^{m} \hat{q}^{n-k} \nonumber\\
\end{align}
Explicitly,
\begin{equation}\label{tsum}
\mathcal{\hat{T}}_{\Psi} = \sum_{m,n}C_{m,n}\hat{T}_{m,n}
\end{equation}
where the $C_{m,n}$'s are just constant $2 \times 2$ matrices.
Also, since the $\sigma_{j}$'s form a complete linearly independent set of $2\times 2$ matrices, any $2 \times 2$ matrix can then be written as a linear sum of the $\sigma_{j}$'s. Specifically, we can write $C_{m,n} = \sum_{j=0}^{3}\gamma_{j}^{m,n}\sigma_{j}$ where $\gamma_{j}^{m,n}$ are just unknown scalars. Substituting Eq (\ref{tsum}) into Eq (\ref{ccr}), we have
\begin{align}\label{frob}
\left[\mathcal{\hat{H}}_{\Psi}, \mathcal{\hat{T}}_{\Psi}\right]\Psi &= \sum_{m,n}\left(c\left[A \hat{p}, C_{m,n} \hat{T}_{m,n}\right] + m_{0}c^{2} \left[B, C_{m,n} \hat{T}_{m,n} \right] \right)\Psi \nonumber \\
&= \sum_{m,n} \left( c\left( A C_{m,n} \hat{p}\hat{T}_{m,n} - C_{m,n} A \hat{T}_{m,n}\hat{p} \right)+ m_{0}c^{2} \left[B, C_{m,n} \right]\hat{T}_{m,n} \right)\Psi \nonumber\\
&= \sum_{m,n} \left( c [A, C_{m,n}] \hat{T}_{m+1,n} - \frac{i\hbar c n}{2} \{A, C_{m,n}\} \hat{T}_{m,n-1}+ m_{0}c^{2} \left[B, C_{m,n} \right]\hat{T}_{m,n} \right)\Psi \nonumber\\
&= \sum_{m,n} \left( c [A, C_{m-1,n}]  - \frac{i\hbar c (n+1)}{2} \{A, C_{m,n+1}\} + m_{0}c^{2} \left[B, C_{m,n} \right] \right)\hat{T}_{m,n} \Psi \nonumber\\
&= \sum_{m,n} i\hbar \sigma_{0} \delta_{m,0}\delta_{n,0} \hat{T}_{m,n} \Psi
\end{align}
where in the fourth line, we shifted indices for each term and in the last line, we assert the equality of the right hand side of Eq (\ref{ccr}). Then, the coefficients of the $\hat{T}_{m,n}$'s should be equal. Additionally, since the $\sigma_{j}$'s are linearly independent, their coefficients should also be equal. These would restrict the values of the coefficients $\gamma_{j}^{m,n}$'s. 

Strictly speaking, the coefficients $\gamma_{j}^{m,0}$ would remain arbitrary. However, we wish to take as many coefficients as possible to vanish. That is, we wish to take the minimal solution so that we can set $\gamma_{j}^{m,0}=0$. It can be shown that the only non-vanishing coefficients are $\gamma_{j=\{1,2,3\}}^{0,1} = -\alpha_{j}/c$ and $\gamma_{j=\{1,2,3\}}^{-1,1} = -m_{0}\beta_{j}$.
Explicitly, we then have
\begin{align}\label{conjt1}
\mathcal{\hat{T}}_{\Psi} &= C_{-1,1}\hat{T}_{-1,1} + C_{0,1}\hat{T}_{0,1} \nonumber\\
&= -m_{0}(\beta_{1}\sigma_{1}+\beta_{2}\sigma_{2}+\beta_{3}\sigma_{3}) \hat{T}_{-1,1} - \frac{1}{c}(\alpha_{1}\sigma_{1}+\alpha_{2}\sigma_{2}+\alpha_{3}\sigma_{3})\hat{T}_{0,1} \nonumber\\
&= -m_{0} B \hat{T}_{-1,1} - \frac{1}{c} A \hat{T}_{0,1} \nonumber\\
\end{align}
so that in momentum representation (specifically, $\Psi - p$ representation), its action on a column matrix $\Psi(p)$ is
\begin{equation}\label{conjt1p}
\left(\mathcal{\hat{T}}_{\Psi}\Psi \right)(p) = -m_{0} B \frac{i\hbar}{2}\left( \frac{1}{p}\frac{\partial}{\partial p}\Psi (p) + \frac{\partial}{\partial p}\left(\frac{1}{p}\Psi (p)\right)\right) -\frac{1}{c}A i\hbar\frac{\partial}{\partial p}\Psi(p)
\end{equation}

Much like in the spin-$0$ case, we transform our operators to the $\Phi - p$ representation. This transformation is made through $\mathcal{\hat{T}}_{\Phi}\Phi = U\mathcal{\hat{T}}_{\Psi}U^{-1}\Phi$ for some arbitrary column matrix $\Phi$. The unitary transformation $U$ diagonalizes the Hamiltonian so that $\mathcal{\hat{H}}_{\Phi}\Phi = U\mathcal{\hat{H}}_{\Psi}U^{-1}\Phi = E_{p}\sigma_{3}\Phi$. Its eigenstates are then explicitly separated into positive and negative states. This makes one-particle analysis more straightforward. Explicitly, $U$ is given by Eq (\ref{utrans}) so that we can calculate
\begin{align}
U\mathcal{\hat{T}}_{\Psi}U^{-1}\Phi &= -m_{0} U B \hat{T}_{-1,1}U^{-1}\Phi - \frac{1}{c} UA \hat{T}_{0,1}U^{-1}\Phi \nonumber\\
&= -m_{0} U B U^{-1}\hat{T}_{-1,1}\Phi - \frac{1}{c} UAU^{-1} \hat{T}_{0,1}U^{-1}\Phi - i\hbar U \left(\frac{A}{c} + \frac{m_{0}B}{p}\right)\frac{\partial U^{-1}}{\partial p}\Phi \nonumber\\
&= -m_{0}U B U^{-1} \left(\frac{i\hbar}{p}\frac{\partial \Phi}{\partial p} -\frac{i\hbar}{2p^{2}}\Phi \right) - \frac{1}{c} U A U^{-1} i\hbar \frac{\partial\Phi }{\partial p} - \frac{i\hbar}{pc^{2}}U \mathcal{\hat{H}}_{\Psi}\frac{\partial U^{-1}}{\partial p}\Phi  \nonumber\\ 
&= - \frac{i\hbar}{pc^{2}}U\mathcal{\hat{H}}_{\Psi}U^{-1}\frac{\partial \Phi}{\partial p} + i\hbar U \left(\frac{m_{0}}{2p^{2}}BU^{-1} - \frac{1}{pc^{2}}\mathcal{\hat{H}}_{\Psi}\frac{\partial U^{-1}}{\partial p} \right)\Phi \nonumber\\
&= - \frac{i\hbar}{pc^{2}} E_{p} \sigma_{3} \frac{\partial \Phi}{\partial p} + i\hbar \frac{m_{0}^{2}c^{2}}{2p^{2}E_{p}}\sigma_{3}\Phi - \frac{(\alpha_{1}\beta_{2}-\alpha_{2}\beta_{1})}{(E_{p}+pc\alpha_{3}+m_{0}c^{2}\beta_{3})}\frac{\hbar m_{0}c}{2p}\Phi \nonumber\\
\end{align}
where we can let
\begin{align}
\mathcal{\hat{T}}\Phi &= - \frac{i\hbar}{pc^{2}} E_{p} \sigma_{3} \frac{\partial \Phi}{\partial p} + i\hbar \frac{m_{0}^{2}c^{2}}{2p^{2}E_{p}}\sigma_{3}\Phi \nonumber\\
&= -\frac{i\hbar}{2E_{p}}\left(\left(2p+2\frac{m_{0}^{2}c^{2}}{p}\right)\sigma_{3}\frac{\partial \Phi}{\partial p} - \frac{m_{0}^{2}c^{2}}{p^{2}}\sigma_{3}\Phi \right)\nonumber\\
&= -\frac{1}{2E_{p}}\sigma_{3} \left(i\hbar p\frac{\partial \Phi}{\partial p} + \frac{i\hbar}{2}\Phi\right) - \frac{1}{2E_{p}}(p^{2}+2m_{0}^{2}c^{2})\sigma_{3}\left(i\hbar \frac{1}{p}\frac{\partial \Phi}{\partial p} - \frac{i\hbar}{2p^{2}}\Phi\right) \nonumber\\
&= -\frac{1}{2E_{p}}\sigma_{3} \hat{T}_{1,1}\Phi - \left( \frac{m_{0}^{2}c^{2}}{2E_{p}} + \frac{E_{p}}{2c^{2}} \right)\sigma_{3}\hat{T}_{-1,1}\Phi \nonumber\\
\mathcal{\hat{T}}_{0}\Phi &= - \frac{(\alpha_{1}\beta_{2}-\alpha_{2}\beta_{1})}{(E_{p}+pc\alpha_{3}+m_{0}c^{2}\beta_{3})}\frac{\hbar m_{0}c}{2p}\Phi \nonumber\\
&= - \frac{(\alpha_{1}\beta_{2}-\alpha_{2}\beta_{1})}{(E_{p}/c^{2}+p\alpha_{3}/c+m_{0}\beta_{3})}\frac{\hbar m_{0}}{2pc}\Phi \nonumber\\
\end{align}
We see that $\mathcal{\hat{T}}$ has the same action for the spin-$0$ case. It was shown that in the non-relativistic limit, $\mathcal{\hat{T}}$ reduces to $-m_{0}\hat{T}_{-1,1}$ which is just the TOA operator for the free non-relativistic particle. Also in the non-relativistic limit, it can be shown that $\mathcal{\hat{T}}_{0}$ vanishes.

\section{Imposing Parity Inversion Symmetry}
\label{symm}
Consider the one-particle parity operator $\Pi$, whose action on a column matrix $\Phi(p)$ in $\Phi - p$ representation is $\Pi\Phi(p) = \Phi(-p)$. This allows us to calculate the following
\begin{align}
\Pi\mathcal{\hat{T}}\Phi(p,t) &= -\frac{i\hbar}{pc^{2}} E_{p} \sigma_{3} \frac{\partial \Phi(-p,t)}{\partial p} + i\hbar \frac{m_{0}^{2}c^{2}}{2p^{2}E_{p}}\sigma_{3}\Phi(-p,t) \nonumber\\
\mathcal{\hat{T}}\Pi\Phi(p,t) &= -\frac{i\hbar}{pc^{2}} E_{p} \sigma_{3} \frac{\partial \Phi(-p,t)}{\partial p} + i\hbar \frac{m_{0}^{2}c^{2}}{2p^{2}E_{p}}\sigma_{3}\Phi(-p,t) \nonumber\\
\left[\Pi, \mathcal{\hat{T}}\right]\Phi(p,t) &= 0 \nonumber\\
\end{align}
where, $df(x)/dx|_{x=-p} = - df(-p)/dp$. This implies that $\mathcal{\hat{T}}$ is invariant under parity inversions, i.e. $\Pi\mathcal{\hat{T}}\Pi^{-1} = \mathcal{\hat{T}}$. Similarly,
\begin{align}
\Pi\mathcal{\hat{T}}_{0}\Phi(p,t) &= \frac{(\alpha_{1}\beta_{2}-\alpha_{2}\beta_{1})}{(E_{p}-pc\alpha_{3}+m_{0}c^{2}\beta_{3})}\frac{\hbar m_{0}c}{2p}\Phi(-p,t)\nonumber\\
\mathcal{\hat{T}}_{0}\Pi\Phi(p,t) &= - \frac{(\alpha_{1}\beta_{2}-\alpha_{2}\beta_{1})}{(E_{p}+pc\alpha_{3}+m_{0}c^{2}\beta_{3})}\frac{\hbar m_{0}c}{2p}\Phi(-p,t)\nonumber\\
\left[\Pi, \mathcal{\hat{T}}_{0}\right]\Phi(p,t) &= \frac{2(E_{p}+m_{0}c^{2}\beta_{3})(\alpha_{1}\beta_{2}-\alpha_{2}\beta_{1})}{(E_{p}+m_{0}c^{2}\beta_{3})^{2}-(pc\alpha_{3})^{2}}\frac{\hbar m_{0}c}{2p}\Phi(-p,t)\nonumber\\
\end{align}
We see then that $\mathcal{\hat{T}}_{0}$ is not invariant under parity inversions. 

If we impose this symmetry on our TOA operator for free spin-$1/2$ particles, then $\mathcal{\hat{T}}_{0}$ must be thrown away. Since it can be shown that $[\mathcal{\hat{H}}_{\Phi},\mathcal{\hat{T}}_{0}] = 0$, we can interpret $\mathcal{\hat{T}}_{0}$ as a term not contributing to the conjugacy of the TOA operator with the Hamiltonian $\mathcal{\hat{H}}_{\Phi}$. Actually, we can add any operator that commutes with $\mathcal{\hat{H}}_{\Phi}$ to one which is conjugate to $\mathcal{\hat{H}}_{\Phi}$, suggesting that there are many solutions to Eq (\ref{ccr}). These operators then represent the many characteristics, roles, and physical interpretations of a time observable \cite{gal6, gal7}. We are certain that the remaining term $\mathcal{\hat{T}}$ represents a TOA observable since its action is the same with the TOA operator from \cite{kgrqtoa} and since the time evolution of states, Eq (\ref{tevol}), is similar to that of the Klein-Gordon case \cite{kgrqtoa,art,book}, the dynamic behavior of the eigenfunctions would also be similar to the Klein-Gordon case \cite{kgrqtoa}. 

As an example, the time evolution of the probability density $\rho = \Psi^{\dagger}\Psi = \Phi^{\dagger}\Phi$ (in configuration space) of a typical eigenfunction $\Phi_{\lambda, \tau}^{(\pm)}$ corresponding to the eigenvalue $\tau = 1.0$ is given in the contour plots \cite{kgrqtoa}.
\begin{figure}[!htb]
\centering
\includegraphics[width=1\textwidth, height=0.415\textwidth]{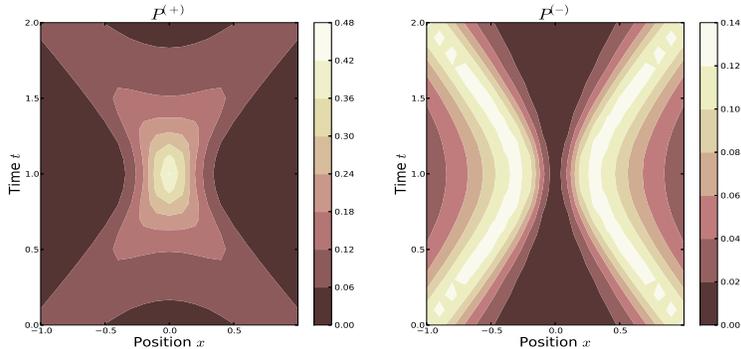}
\caption{Time Evolution of the Probability Density with $\tau = 1.0$}
\label{pdense10}
\end{figure}
It can be seen that in the $P^{(+)}$ contour plot, the probability density peaks about the origin at the time of the eigenvalue. Whereas in the $P^{(-)}$ contour plot, the two peaks are closest to the origin at the time of the eigenvalue, but with the probability always vanishing at the origin. We call the states with similar behaviors as demonstrated in $P^{(+)}$ as a non-nodal eigenfunction $\Phi_{\lambda, \tau}^{(+)}$, while the states with similar behaviors as in $P^{(-)}$ as a nodal eigenfunction $\Phi_{\lambda, \tau}^{(-)}$. The localization of the states at the origin at their corresponding eigenvalues is then interpreted as the particle arriving at the origin at the time of the eigenvalue. That is, the eigenfunctions represent states of definite arrival time at the origin. The specific difference would be the inner products defined on the vector spaces on which they act. Hence, a one-particle TOA operator for a free spin-$1/2$ relativistic particle is given by $\mathcal{\hat{T}}$ whose action is the same as the TOA operator for the spin-$0$ case.

\section{Conclusion} 
\label{conc}

We have constructed the minimal operator solution $\mathcal{\hat{T}}_{\Psi}$ that is conjugate with the Hamiltonian for a relativistic spin-$1/2$ particle. It is a true one-particle operator since it does not map positive states into negative states and vice-versa. That is, it is diagonal in the $\Phi$-representation. We see that the action of the said operator $\mathcal{\hat{T}}_{\Phi} = U\mathcal{\hat{T}}_{\Psi}U^{-1} = \mathcal{\hat{T}} + \mathcal{\hat{T}}_{0}$ is the same as the one derived from \cite{kgrqtoa} apart from an additional operator term $\mathcal{\hat{T}}_{0}$ which commutes with the Hamiltonian. But upon imposing that a TOA operator should be invariant upon parity inversion, we see that $\mathcal{\hat{T}}_{0}$ breaks this symmetry. We can throw away this term so that our relativistic one-particle TOA operator for a free spin-$1/2$ particle is $\mathcal{\hat{T}}$. (Any other implications of $\mathcal{\hat{T}}_{0}$, if any, may be studied later.) Since the calculated actions of our TOA operators for the spin-$0$ and spin-$1/2$ free particles are the same, and the time evolution of states are the same, we expect similar properties. As an example, they should both reduce to the non-relativistic TOA operator $-m_{0}\hat{T}_{-1,1}$, and the form and characteristics of their eigenfunctions must be the same. The difference would be in the inner products defined on the Hilbert spaces they are acting upon. Then in this relativistic quantum regime, time is an observable represented by an operator. More importantly, it was constructed by imposing a quantum principle and hence, without resorting to quantization.

One of the motivations of studying time in relativistic quantum mechanics is to lessen the inconsistencies between the notions of time in special relativity and quantum mechanics. The hope is that, from these studies, we may gain further insight that can help us mend the marriage of general relativity and quantum mechanics. The reconciliation of which is very much harder. Perhaps that insight is supraquantization. Consider the standard canonical quantum general relativity approach for example, spacetime is split into space and time so that the classical equations describing the system can be cast into Hamiltonian form \cite{qgbook1, qgbook2}. This Hamiltonian however, is actually constrained to vanish via the equations of motion. The quantum version of this is supposedly to impose the vanishing of the quantized Hamiltonian as a condition on the physical states. This results into the Wheeler-DeWitt equation - similar in form to the Schrodinger equation but with the time derivative vanishing or 'missing'. This vanishing leads to one facet of the problem of time in quantum gravity \cite{qgart} (Is the universe 'frozen'?, Is time an illusion?). Perhaps this is where supraquantization can come into play. Perhaps there is a more 'fundamental' way to construct operators, like the Hamiltonian stated earlier, by using some relevant quantum first principle(s) so that we won't need quantization. Of course, even determining the appropriate first principles would require further studies to plant these speculations on firmer and more rigorous grounds. Nevertheless, these studies may lead us closer on formulating a theory of quantum spacetime.

\appendix
\section{Calculation Details of the Construction of $\mathcal{\hat{T}}_{\Psi}$}
\label{constrdet}
We consider the commutator between the position and momentum operators $\hat{q}$ and $\hat{p}$, respectively, $\left[\hat{q},\hat{p}\right] = i\hbar$. We calculate the following:
\begin{align}
\left[\hat{p},\hat{q}^{k}\right] &=\left[\hat{p},\hat{q}^{k-1}\hat{q}\right] \nonumber\\
&=\hat{q}^{k-1}\left[\hat{p},\hat{q}\right]+\left[\hat{p},\hat{q}^{k-1}\right]\hat{q} \nonumber\\
&=-i\hbar\hat{q}^{k-1}+\left[\hat{p},\hat{q}^{k-2}\hat{q}\right]\hat{q} \nonumber\\
&=-i\hbar\hat{q}^{k-1}+(\hat{q}^{k-2}\left[\hat{p},\hat{q}\right]+\left[\hat{p},\hat{q}^{k-2}\right]\hat{q})\hat{q} \nonumber\\
&=-2i\hbar\hat{q}^{k-1}+\left[\hat{p},\hat{q}^{k-2}\right]\hat{q}^{2} \nonumber\\
& \;\;\vdots \nonumber\\
&=-i\hbar k\hat{q}^{k-1} \nonumber\\
\end{align}
Now, consider the complete and linearly independent set of Bender-Dunne operators denoted by $\hat{T}_{m,n}$.
\begin{align}
\hat{T}_{m,n} &=\frac{1}{2^n}\sum_{k=0}^{\infty} \frac{n!}{k!(n-k)!} \hat{q}^{k} \hat{p}^{m} \hat{q}^{n-k} \nonumber\\
&=\frac{1}{2^n}\sum_{k=0}^{\infty} \frac{n!}{k!(n-k)!} \hat{q}^{n-k} \hat{p}^{m} \hat{q}^{k} \nonumber\\
\end{align}
and calculate the following:
\begin{align}
\hat{p}\hat{T}_{m,n} &= \frac{1}{2^n}\sum_{k=0}^{\infty} \frac{n!}{k!(n-k)!} (\hat{p}\hat{q}^{k}) \hat{p}^{m} \hat{q}^{n-k} \nonumber\\
&= \frac{1}{2^n}\sum_{k=0}^{\infty} \frac{n!}{k!(n-k)!} (\hat{q}^{k}\hat{p} - i\hbar k\hat{q}^{k-1}) \hat{p}^{m} \hat{q}^{n-k} \nonumber\\
&= \hat{T}_{m+1,n} - \frac{i\hbar n}{2}\hat{T}_{m,n-1} \nonumber\\
\hat{T}_{m,n}\hat{p} &= \frac{1}{2^n}\sum_{k=0}^{\infty} \frac{n!}{k!(n-k)!} \hat{q}^{n-k} \hat{p}^{m} (\hat{q}^{k}\hat{p}) \nonumber\\
&= \frac{1}{2^n}\sum_{k=0}^{\infty} \frac{n!}{k!(n-k)!}\hat{q}^{n-k}\hat{p}^{m} (\hat{p}\hat{q}^{k} + i\hbar k\hat{q}^{k-1} ) \nonumber\\
&= \hat{T}_{m+1,n} + \frac{i\hbar n}{2}\hat{T}_{m,n-1} \nonumber\\
\end{align}

We also consider a pair of $2\times 2$ matrices $A=\sum_{j=0}^{3}\alpha_{j}\sigma_{j}$ and $B=\sum_{j=0}^{3}\beta_{j}\sigma_{j}$ satisfying the anti-commutation relation $\{A,B\} = 0$ and $A^{2}=B^{2}=\sigma_{0}$ where the set of $\sigma_{j}$'s form a complete linearly independent set of $2\times 2$ matrices, denoted by
\begin{displaymath}
\sigma_{0} = 
\begin{pmatrix}
 1 & 0\\
 0 & 1 \\
\end{pmatrix},
\sigma_{1} = 
\begin{pmatrix}
 0 & 1\\
 1 & 0 \\
\end{pmatrix},
\sigma_{2} = 
\begin{pmatrix}
 0 & -i\\
 i & 0 \\
\end{pmatrix},
\sigma_{3} = 
\begin{pmatrix}
 1 & 0\\
 0 & -1 \\
\end{pmatrix}
\end{displaymath}
Note that $\sigma_{0}$ is just the identity matrix. Some properties of these matrices include $\sigma_{x}\sigma_{y}=i\sigma_{z}$ where, $(x,y,z)$ are just cyclic permutations of $(1,2,3)$, and $\sigma_{j}\sigma_{j} = \sigma_{0}$ for $j=0,1,2,3$. The condition that $A$ and $B$ anti-commute and that both of their squares equal to the identity matrix imply that we should have the following conditions $\alpha_{0}=\beta_{0}=0$, $\sum_{j=1}^{3}\alpha_{j}\beta_{j} = 0$, and $\sum_{j=1}^{3}\alpha_{j}^{2} = \sum_{j=1}^{3}\beta_{j}^{2} = 1$. We also restrict the $\alpha_{j}$'s and $\beta_{j}$'s to be real so that $A$ and $B$ are self-adjoint. The Hamiltonian $\mathcal{\hat{H}}_{\Psi} = cA\hat{p} + m_{0}c^{2}B$ is then also self-adjoint. Using the properties of the $\sigma_{j}$'s, we can calculate some relevant commutation and anti-commutation relations of $A$ and $B$ with $C_{m,n} = \sum_{j=0}^{3}\gamma_{j}^{m,n}\sigma_{j}$. Specifically,
\begin{align}
[A,C_{m-1,n}] &= \gamma_{1}^{m-1,n}[A,\sigma_{1}] + \gamma_{2}^{m-1,n}[A,\sigma_{2}] + \gamma_{3}^{m-1,n}[A,\sigma_{3}] \nonumber\\
&= \gamma_{1}^{m-1,n}(-2i\alpha_{2}\sigma_{3} + 2i\alpha_{3}\sigma_{2}) + \gamma_{2}^{m-1,n}(2i\alpha_{1}\sigma_{3} - 2i\alpha_{3}\sigma_{1}) \nonumber\\
& \;\;\;\;\; + \gamma_{3}^{m-1,n}(-2i\alpha_{1}\sigma_{2} + 2i\alpha_{2}\sigma_{1}) \nonumber\\
&= 2i\left( (\gamma_{3}^{m-1,n} \alpha_{2} - \gamma_{2}^{m-1,n}\alpha_{3})\sigma_{1} + (\gamma_{1}^{m-1,n} \alpha_{3} - \gamma_{3}^{m-1,n}\alpha_{1})\sigma_{2} \right. \nonumber\\
& \;\;\;\;\; \left. + (\gamma_{2}^{m-1,n} \alpha_{1} - \gamma_{1}^{m-1,n}\alpha_{2})\sigma_{3} \right) \nonumber\\
[B,C_{m,n}] &= 2i\left( (\gamma_{3}^{m,n} \beta_{2} - \gamma_{2}^{m,n}\beta_{3})\sigma_{1} + (\gamma_{1}^{m,n} \beta_{3} - \gamma_{3}^{m,n}\beta_{1})\sigma_{2} \right. \nonumber\\
& \;\;\;\;\; \left. + (\gamma_{2}^{m,n} \beta_{1} - \gamma_{1}^{m,n}\beta_{2})\sigma_{3} \right) \nonumber\\
\{A, C_{m,n+1}\} &= \gamma_{0}^{m,n+1}\{A,\sigma_{0}\} + \gamma_{1}^{m,n+1}\{A,\sigma_{1}\} + \gamma_{2}^{m,n+1}\{A,\sigma_{2}\} + \gamma_{3}^{m,n+1}\{A,\sigma_{3}\} \nonumber\\
&= 2\gamma_{0}^{m,n+1}(\alpha_{1}\sigma_{1} + \alpha_{2}\sigma_{2} + \alpha_{3}\sigma_{3}) + 2\gamma_{1}^{m,n+1}\alpha_{1}\sigma_{0} \nonumber\\
& \;\;\;\;\; + 2\gamma_{2}^{m,n+1}\alpha_{2}\sigma_{0} + 2\gamma_{3}^{m,n+1}\alpha_{3}\sigma_{0} \nonumber\\
&= 2(\gamma_{1}^{m,n+1}\alpha_{1} + \gamma_{2}^{m,n+1}\alpha_{2} + \gamma_{3}^{m,n+1}\alpha_{3}) \sigma_{0} \nonumber\\
& \;\;\;\;\; + 2\gamma_{0}^{m,n+1}\alpha_{1}\sigma_{1} + 2\gamma_{0}^{m,n+1}\alpha_{2}\sigma_{2} + 2\gamma_{0}^{m,n+1}\alpha_{3}\sigma_{3}\nonumber\\
\end{align}

From Eq (\ref{frob}), we then arrive to four equations restricting the values of the $\gamma_{j}^{m,n}$'s.
\begin{align}
-\frac{i\hbar c}{2}(n+1)2\left(\gamma_{1}^{m,n+1}\alpha_{1}+\gamma_{2}^{m,n+1}\alpha_{2}+\gamma_{3}^{m,n+1}\alpha_{3}\right) &= i\hbar \delta_{m,0}\delta_{n,0} \label{eq1} \\
c2i(\gamma_{3}^{m-1,n}\alpha_{2}-\gamma_{2}^{m-1,n}\alpha_{3})-\frac{i\hbar c}{2}(n+1)2\gamma_{0}^{m,n+1}\alpha_{1} \;\;\; & \nonumber\\
+ m_{0}c^{2}2i(\gamma_{3}^{m,n}\beta_{2}-\gamma_{2}^{m,n}\beta_{3}) &= 0 \label{eq2} \\
c2i(\gamma_{1}^{m-1,n}\alpha_{3}-\gamma_{3}^{m-1,n}\alpha_{1})-\frac{i\hbar c}{2}(n+1)2\gamma_{0}^{m,n+1}\alpha_{2} \;\;\; & \nonumber\\
+ m_{0}c^{2}2i(\gamma_{1}^{m,n}\beta_{3}-\gamma_{3}^{m,n}\beta_{1}) &= 0 \label{eq3} \\
c2i(\gamma_{2}^{m-1,n}\alpha_{1}-\gamma_{1}^{m-1,n}\alpha_{2})-\frac{i\hbar c}{2}(n+1)2\gamma_{0}^{m,n+1}\alpha_{3} \;\;\; & \nonumber\\
+ m_{0}c^{2}2i(\gamma_{2}^{m,n}\beta_{1}-\gamma_{1}^{m,n}\beta_{2}) &= 0 \label{eq4} \\
\nonumber
\end{align}
Using Eqs (\ref{eq2}-\ref{eq4}), to cancel their respective third terms, we arrive at
\begin{align}\label{eq5}
\gamma_{1}^{m-1,n}(\beta_{2}\alpha_{3}-\beta_{3}\alpha_{2}) &+ \gamma_{2}^{m-1,n}(\beta_{3}\alpha_{1}-\beta_{1}\alpha_{3}) + \gamma_{3}^{m-1,n}(\beta_{1}\alpha_{2}-\beta_{2}\alpha_{1}) \nonumber\\
& - \frac{\hbar}{2}(n+1)\gamma_{0}^{m,n+1}\cancelto{0}{(\beta_{1}\alpha_{1}+\beta_{2}\alpha_{2} + \beta_{3}\alpha_{3})} = 0 \nonumber\\
\end{align}
Using again Eqs (\ref{eq2}-\ref{eq4}), this time to cancel their respective first terms, we arrive at 
\begin{align}\label{eq6}
- m_{0}c( \gamma_{1}^{m,n}(\beta_{2}\alpha_{3}-\beta_{3}\alpha_{2}) &+ \gamma_{2}^{m,n}(\beta_{3}\alpha_{1}-\beta_{1}\alpha_{3}) + \gamma_{3}^{m,n}(\beta_{1}\alpha_{2}-\beta_{2}\alpha_{1}) ) \nonumber\\
& - \frac{\hbar}{2}(n+1)\gamma_{0}^{m,n+1}\cancelto{1}{(\alpha_{1}^{2}+\alpha_{2}^{2} + \alpha_{3}^{2})} = 0 \nonumber\\
\end{align}
so that Eq (\ref{eq5}) and Eq (\ref{eq6}) imply that $\gamma_{0}^{m,n \neq 0} = 0$. For $n \neq 0$, Eqs (\ref{eq1}-\ref{eq4}) can then be written as
\begin{align}
-c n\left(\gamma_{1}^{m,n}\alpha_{1}+\gamma_{2}^{m,n}\alpha_{2}+\gamma_{3}^{m,n}\alpha_{3}\right) &= \delta_{m,0}\delta_{n,1} \label{eq7} \\
\gamma_{3}^{m-1,n}\alpha_{2} + m_{0} c \gamma_{3}^{m,n} \beta_{2} &= \gamma_{2}^{m-1,n}\alpha_{3} + m_{0} c \gamma_{2}^{m,n} \beta_{3} \label{eq8} \\
\gamma_{1}^{m-1,n}\alpha_{3} + m_{0} c \gamma_{1}^{m,n} \beta_{3} &= \gamma_{3}^{m-1,n}\alpha_{1} + m_{0} c \gamma_{3}^{m,n} \beta_{1} \label{eq9} \\
\gamma_{2}^{m-1,n}\alpha_{1} + m_{0} c \gamma_{2}^{m,n} \beta_{1} &= \gamma_{1}^{m-1,n}\alpha_{2} + m_{0} c \gamma_{1}^{m,n} \beta_{2} \label{eq10} \\ \nonumber
\end{align}
Appropriately combining Eqs (\ref{eq8}) and (\ref{eq9}), (\ref{eq8}) and (\ref{eq10}), and (\ref{eq9}) and (\ref{eq10}), respectively, we arrive at 
\begin{align}
\alpha_{3}(\gamma_{1}^{m-1,n}\alpha_{1}+&\gamma_{2}^{m-1,n}\alpha_{2}) + m_{0}c (\gamma_{1}^{m,n}\alpha_{1}+\gamma_{2}^{m,n}\alpha_{2}) \beta_{3} \nonumber\\
& = \gamma_{3}^{m-1,n} \cancelto{1-\alpha_{3}^{2}}{(\alpha_{1}^{2}+\alpha_{2}^{2})} + m_{0}c \gamma_{3}^{m,n} \cancelto{-\alpha_{3} \beta_{3}}{(\alpha_{1}\beta_{1}+\alpha_{2}\beta_{2})} \nonumber\\
\alpha_{2}(\gamma_{1}^{m-1,n}\alpha_{1}+&\gamma_{3}^{m-1,n}\alpha_{3}) + m_{0}c (\gamma_{1}^{m,n}\alpha_{1}+\gamma_{3}^{m,n}\alpha_{3}) \beta_{2} \nonumber\\
& = \gamma_{2}^{m-1,n} \cancelto{1-\alpha_{2}^{2}}{(\alpha_{1}^{2}+\alpha_{3}^{2})} + m_{0}c \gamma_{2}^{m,n} \cancelto{-\alpha_{2} \beta_{2}}{(\alpha_{1}\beta_{1}+\alpha_{3}\beta_{3})} \nonumber\\
\alpha_{1}(\gamma_{2}^{m-1,n}\alpha_{2}+&\gamma_{3}^{m-1,n}\alpha_{3}) + m_{0}c (\gamma_{2}^{m,n}\alpha_{2}+\gamma_{3}^{m,n}\alpha_{3}) \beta_{1} \nonumber\\
& = \gamma_{1}^{m-1,n} \cancelto{1-\alpha_{1}^{2}}{(\alpha_{2}^{2}+\alpha_{3}^{2})} + m_{0}c \gamma_{1}^{m,n} \cancelto{-\alpha_{1} \beta_{1}}{(\alpha_{2}\beta_{2}+\alpha_{3}\beta_{3})} \nonumber\\
\nonumber
\end{align}
and using Eq (\ref{eq7}), we get $n\gamma_{j=\{1,2,3\}}^{m,n\neq 0} = -\delta_{m,0}\delta_{n,1}\alpha_{j}/c - m_{0}\delta_{m,-1}\delta_{n,1}\beta_{j}$. Note however, that there are some arbitrary coefficients $\gamma_{j=\{0,1,2,3\}}^{m,0}$ which do not necessarily vanish. We wish to take the minimal solution so that we take to zero as many $\gamma_{j}^{m,n}$'s as possible. That is, we set the arbitrary coefficients $\gamma_{j}^{m,0} = 0$. The only non-vanishing coefficients are then $\gamma_{j=\{1,2,3\}}^{0,1} = -\alpha_{j}/c$ and $\gamma_{j=\{1,2,3\}}^{-1,1} = -m_{0}\beta_{j}$.

\section{Calculation Details of the Transformation $\mathcal{\hat{T}}_{\Phi} = U\mathcal{\hat{T}}_{\Psi}U^{-1}$}
\label{transdet}
We first take the derivative of Eq (\ref{utrans})
\begin{align}
\frac{\partial U^{-1}}{\partial p} &= \frac{(E_{p}+\alpha_{3}pc + \beta_{3}m_{0}c^{2})^{-1/2}}{\sqrt{2}}\left(\frac{pc^{2}}{2E_{p}^{3/2}}\sigma_{3}+\frac{cA}{E_{p}^{1/2}}-\frac{pc^{2}}{2E_{p}^{5/2}}\mathcal{\hat{H}}_{\Psi} \right) \nonumber\\
& \;\;\; -\frac{(E_{p}+\alpha_{3}pc + \beta_{3}m_{0}c^{2})^{-3/2}}{2\sqrt{2}}\left( \frac{pc^{2}}{E_{p}} + \alpha_{3}c \right) \frac{E_{p}\sigma_{3}+ \mathcal{\hat{H}}_{\Psi}}{ E_{p}^{1/2} }\nonumber\\
&= \frac{(E_{p}+\alpha_{3}pc + \beta_{3}m_{0}c^{2})^{-3/2}}{2E_{p}\sqrt{2E_{p}}}\left[ -(pc^{2}+cE_{p}\alpha_{3})(E_{p}\sigma_{3}+ \mathcal{\hat{H}}_{\Psi}) \right.\nonumber\\
& \;\;\; +\left. \left(pc^{2}\sigma_{3}+2cE_{p}A-\frac{pc^{2}}{E_{p}}\mathcal{\hat{H}}_{\Psi}\right)(E_{p}+\alpha_{3}pc + \beta_{3}m_{0}c^{2}) \right]\nonumber\\
&= \frac{(E_{p}+\alpha_{3}pc + \beta_{3}m_{0}c^{2})^{-3/2}}{2E_{p}\sqrt{2E_{p}}}\left[ (p^{2}c^{3}\alpha_{3}-cE_{p}^{2}\alpha_{3} + pm_{0}c^{4}\beta_{3})\sigma_{3} \right.\nonumber\\
& \;\;\; \left. +2(cE_{p}^{2}+pc^{2}E_{p}\alpha_{3}+m_{0}c^{3}E_{p}\beta_{3})A -\left(2pc^{2}+\frac{p^{2}c^{3}\alpha_{3}}{E_{p}}+cE_{p}\alpha_{3}+\frac{pm_{0}c^{4}\beta_{3}}{E_{p}}\right)\mathcal{\hat{H}}_{\Psi} \right]\nonumber\\
\end{align}
Remembering that $\mathcal{\hat{H}}_{\Psi}^{2} = E_{p}^{2}\sigma_{0} = (p^{2}c^{2}+m_{0}^{2}c^{4})\sigma_{0}$, we calculate
\begin{align}
\frac{\mathcal{\hat{H}}_{\Psi}}{pc^{2}}\frac{\partial U^{-1}}{\partial p} &= \frac{(E_{p}+\alpha_{3}pc + \beta_{3}m_{0}c^{2})^{-\frac{3}{2}}}{2E_{p}\sqrt{2E_{p}}}\left[ \left(m_{0}c^{2}\beta_{3}-\frac{m_{0}^{2}c^{3}\alpha_{3}}{p}\right)\mathcal{\hat{H}}_{\Psi}\sigma_{3} \right.\nonumber\\
& \;\;\;  +2\left(\frac{E_{p}^{2}}{pc}+E_{p}\alpha_{3}+\frac{m_{0}cE_{p}\beta_{3}}{p}\right)(pc\sigma_{0}+m_{0}c^{2}BA) \nonumber\\
& \;\;\; - \left. \left(2E_{p}^{2}+pcE_{p}\alpha_{3}+\frac{E_{p}^{3}\alpha_{3}}{pc}+m_{0}c^{2}E_{p}\beta_{3}\right)\sigma_{0} \right] \nonumber\\
&= \frac{(E_{p}+\alpha_{3}pc + \beta_{3}m_{0}c^{2})^{-\frac{3}{2}}}{2E_{p}\sqrt{2E_{p}}}\left[ \left(m_{0}c^{2}\beta_{3}-\frac{m_{0}^{2}c^{3}\alpha_{3}}{p}\right)  (pcA\sigma_{3}+m_{0}c^{2}B\sigma_{3}) \right.\nonumber\\
& \;\;\; \left. +2\left(\frac{m_{0}cE_{p}^{2}}{p}+m_{0}c^{2}E_{p}\alpha_{3}+\frac{m_{0}^{2}c^{3}E_{p}\beta_{3}}{p}\right)BA + \left(m_{0}c^{2}\beta_{3}-\frac{m_{0}^{2}c^{3}\alpha_{3}}{p}\right)E_{p}\sigma_{0} \right]\nonumber\\
\end{align}
Afterwards, we also calculate
\begin{align}
\frac{m_{0}BU^{-1}}{2p^{2}} &= \frac{(E_{p}+\alpha_{3}pc + \beta_{3}m_{0}c^{2})^{-\frac{3}{2}}}{2E_{p}\sqrt{2E_{p}}} \frac{m_{0}E_{p}B}{p^{2}}(E_{p}+\alpha_{3}pc + \beta_{3}m_{0}c^{2})(E_{p}\sigma_{3}+\mathcal{\hat{H}}_{\Psi})\nonumber\\
& = \frac{(E_{p}+\alpha_{3}pc + \beta_{3}m_{0}c^{2})^{-\frac{3}{2}}}{2E_{p}\sqrt{2E_{p}}} \left[ \left( \frac{m_{0}E_{p}^{3}}{p^{2}} + \frac{m_{0}cE_{p}^{2}\alpha_{3}}{p} + \frac{m_{0}^{2}c^{2}E_{p}^{2}\beta_{3}}{p^{2}} \right)B\sigma_{3} \right. \nonumber\\
& \;\;\; \left.+ \left( \frac{m_{0}E_{p}^{2}}{p^{2}} + \frac{m_{0}cE_{p}\alpha_{3}}{p} + \frac{m_{0}^{2}c^{2}E_{p}\beta_{3}}{p^{2}}\right)(pcBA+m_{0}c^{2}\sigma_{0}) \right]\nonumber\\
& = \frac{(E_{p}+\alpha_{3}pc + \beta_{3}m_{0}c^{2})^{-\frac{3}{2}}}{2E_{p}\sqrt{2E_{p}}} \left[ \left(\frac{m_{0}cE_{p}^{2}}{p}+m_{0}c^{2}E_{p}\alpha_{3}+\frac{m_{0}^{2}c^{3}E_{p}\beta_{3}}{p}\right)BA \right. \nonumber\\
& \;\;\; + \left( \frac{m_{0}^{2}c^{2}E_{p}^{2}}{p^{2}} + \frac{m_{0}^{2}c^{3}E_{p}\alpha_{3}}{p} + \frac{m_{0}^{3}c^{4}E_{p}\beta_{3}}{p^{2}}\right)\sigma_{0}\nonumber\\
& \;\;\; + \left. \left(\frac{m_{0}E_{p}^{3}}{p^{2}} + pm_{0}c^{3}\alpha_{3} + \frac{m_{0}^{3}c^{5}\alpha_{3}}{p} + m_{0}^{2}c^{4}\beta_{3} + \frac{m_{0}^{4}c^{6}}{p^{2}}\beta_{3} \right)B\sigma_{3} \right]\nonumber\\
\end{align}
so that we can subtract them
\begin{align}
\frac{m_{0}BU^{-1}}{2p^{2}} - \frac{\mathcal{\hat{H}}_{\Psi}}{pc^{2}}\frac{\partial U^{-1}}{\partial p} &= \frac{(E_{p}+\alpha_{3}pc + \beta_{3}m_{0}c^{2})^{-\frac{3}{2}}}{2E_{p}\sqrt{2E_{p}}} \left[ (m_{0}^{2}c^{4}\alpha_{3} - pm_{0}c^{3}\beta_{3})A\sigma_{3} \right. \nonumber\\
& \;\;\; + \left(\frac{m_{0}E_{p}^{3}}{p^{2}} +\left( pm_{0}c^{3} + 2\frac{m_{0}^{3}c^{5}}{p}\right)\alpha_{3} + \frac{m_{0}^{4}c^{6}}{p^{2}}\beta_{3} \right)B\sigma_{3} \nonumber\\
& \;\;\; + \left(\frac{m_{0}^{2}c^{2}E_{p}^{2}}{p^{2}} + 2\frac{m_{0}^{2}c^{3}E_{p}\alpha_{3}}{p} + \left(\frac{m_{0}^{3}c^{4}}{p^{2}}-m_{0}c^{2}\right)E_{p}\beta_{3} \right)\sigma_{0} \nonumber\\
& \;\;\; + \left. \left(\frac{m_{0}cE_{p}^{2}}{p}+m_{0}c^{2}E_{p}\alpha_{3}+\frac{m_{0}^{2}c^{3}E_{p}\beta_{3}}{p}\right) AB \right] \nonumber\\
\end{align}
Lastly, we muliply the result by $U$.
\begin{align}
U\left(\frac{m_{0}BU^{-1}}{2p^{2}}\right. - & \left. \frac{\mathcal{\hat{H}}_{\Psi}}{pc^{2}}\frac{\partial U^{-1}}{\partial p}\right) = \frac{(E_{p}\sigma_{3}+pcA + m_{0}c^{2}B)}{\sqrt{2E_{p}(E_{p}+\alpha_{3}pc+\beta_{3}m_{0}c^{2})}}\left(\frac{m_{0}BU^{-1}}{2p^{2}} - \frac{\mathcal{\hat{H}}_{\Psi}}{pc^{2}}\frac{\partial U^{-1}}{\partial p}\right) \nonumber\\
&= \frac{(E_{p}+\alpha_{3}pc + \beta_{3}m_{0}c^{2})^{-2}}{4 E_{p}^{2}} \left[ (pm_{0}^{2}c^{5}\alpha_{3} - p^{2} m_{0} c^{4} \beta_{3}) A^{2}\sigma_{3} \right. \nonumber\\
& \;\;\; + \left(\frac{m_{0}^{2}c^{2}E_{p}^{3}}{p^{2}} + \frac{m_{0}^{2}c^{3}}{p}( p^{2}c^{2} + 2m_{0}^{2}c^{4})\alpha_{3} + \frac{m_{0}^{5}c^{8}}{p^{2}}\beta_{3} \right) B^{2}\sigma_{3} \nonumber\\
& \;\;\; + \left(\frac{m_{0}^{2}c^{2}E_{p}^{3}}{p^{2}} + 2\frac{m_{0}^{2}c^{3}E_{p}^{2}\alpha_{3}}{p} + \frac{m_{0}}{p^{2}}\left(m_{0}^{2}c^{4}-p^{2}c^{2}\right)E_{p}^{2}\beta_{3} \right)\sigma_{3}\sigma_{0} \nonumber\\
& \;\;\; + \left(\frac{m_{0}^{2}c^{3}E_{p}^{2}}{p} + 2m_{0}^{2}c^{4}E_{p}\alpha_{3} + \left(\frac{m_{0}^{3}c^{5}}{p}-pm_{0}c^{3}\right)E_{p}\beta_{3} \right)A\sigma_{0} \nonumber\\
& \;\;\; + \left(\frac{m_{0}^{2}c^{3}E_{p}^{2}}{p}+m_{0}^{2}c^{4}E_{p}\alpha_{3}+\frac{m_{0}^{3}c^{5}E_{p}\beta_{3}}{p}\right) BAB \nonumber\\
& \;\;\; + (m_{0}^{2}c^{4}E_{p}\alpha_{3} - pm_{0}c^{3}E_{p}\beta_{3})\sigma_{3}A\sigma_{3} \nonumber\\
& \;\;\; + \left(\frac{m_{0}^{3}c^{4}E_{p}^{2}}{p^{2}} + 2\frac{m_{0}^{3}c^{5}E_{p}\alpha_{3}}{p} + \left(\frac{m_{0}^{4}c^{6}}{p^{2}}-m_{0}^{2}c^{4}\right)E_{p}\beta_{3} \right)B\sigma_{0} \nonumber\\
& \;\;\; + \left(m_{0}c^{2}E_{p}^{2}+pm_{0}c^{3}E_{p}\alpha_{3}+m_{0}^{2}c^{4}E_{p}\beta_{3}\right) A^{2}B \nonumber\\
& \;\;\; + \left(\frac{m_{0}E_{p}^{4}}{p^{2}} +\left( pm_{0}c^{3}E_{p} + 2\frac{m_{0}^{3}c^{5}E_{p}}{p}\right)\alpha_{3} + \frac{m_{0}^{4}c^{6}E_{p}}{p^{2}}\beta_{3} \right)\sigma_{3}B\sigma_{3} \nonumber\\
& \;\;\; + \left(\frac{m_{0}cE_{p}^{3}}{p}+m_{0}c^{2}E_{p}^{2}\alpha_{3}+\frac{m_{0}^{2}c^{3}E_{p}^{2}\beta_{3}}{p}\right) \sigma_{3}AB \nonumber\\
& \;\;\; + \left(\frac{m_{0}cE_{p}^{3}}{p} +\left( p^{2}m_{0}c^{4} + 2m_{0}^{3}c^{6}\right)\alpha_{3} + \frac{m_{0}^{4}c^{7}}{p}\beta_{3} \right)AB\sigma_{3} \nonumber\\
& \;\;\; + (m_{0}^{3}c^{6}\alpha_{3} - pm_{0}^{2}c^{5}\beta_{3})BA\sigma_{3}\nonumber\\
&= \frac{(E_{p}+\alpha_{3}pc + \beta_{3}m_{0}c^{2})^{-2}}{4 E_{p}^{2}} \left[ (m_{0}^{2}c^{4}E_{p}\alpha_{3} - pm_{0}c^{3}E_{p}\beta_{3})(A + \sigma_{3}A\sigma_{3}) \right. \nonumber\\
& \;\;\; + \left(\frac{m_{0}E_{p}^{4}}{p^{2}} +\left( pm_{0}c^{3}E_{p} + 2\frac{m_{0}^{3}c^{5}E_{p}}{p}\right)\alpha_{3} + \frac{m_{0}^{4}c^{6}E_{p}}{p^{2}}\beta_{3} \right)(B+\sigma_{3}B\sigma_{3}) \nonumber\\
& \;\;\; + \left( 2\frac{m_{0}^{2}c^{2}E_{p}^{3}}{p^{2}} + 4\frac{m_{0}^{2}c^{3}E_{p}^{2}\alpha_{3}}{p} + 2\frac{m_{0}}{p^{2}}\left(m_{0}^{2}c^{4}-p^{2}c^{2}\right)E_{p}^{2}\beta_{3} \right)\sigma_{3}\nonumber\\
& \;\;\; + \left. \left(\frac{m_{0}cE_{p}^{3}}{p}+m_{0}c^{2}E_{p}^{2}\alpha_{3}+\frac{m_{0}^{2}c^{3}E_{p}^{2}\beta_{3}}{p}\right)( \sigma_{3}AB + AB\sigma_{3} ) \right] \nonumber\\
\end{align}
\begin{align}
\;\;\;\;\;\;\;\;\;\; &= \frac{(E_{p}+\alpha_{3}pc + \beta_{3}m_{0}c^{2})^{-2}}{4 E_{p}^{2}} \left[ \left( 2\frac{m_{0}^{2}c^{2}E_{p}^{3}}{p^{2}} + 4\frac{m_{0}^{2}c^{3}E_{p}^{2}\alpha_{3}}{p} + 4\frac{m_{0}^{3}c^{4}E_{p}^{2}\beta_{3}}{p^{2}} \right. \right.\nonumber\\
& \;\;\;\;\;\;\;\;\; \left. + 2m_{0}^{2}c^{4}E_{p}\alpha_{3}^{2} + 2\frac{m_{0}^{4}c^{6}E_{p}\beta_{3}^{2}}{p^{2}} + 4\frac{m_{0}^{3}c^{5}E_{p}\alpha_{3}\beta_{3}}{p} \right)\sigma_{3} \nonumber\\
& \;\;\;\;\;\;\;\;\; \left. + \left(\frac{m_{0}cE_{p}^{3}}{p}+m_{0}c^{2}E_{p}^{2}\alpha_{3}+\frac{m_{0}^{2}c^{3}E_{p}^{2}\beta_{3}}{p}\right)2i(\alpha_{1}\beta_{2}-\alpha_{2}\beta_{1})\sigma_{0} \right] \nonumber\\
&= \frac{(E_{p}+\alpha_{3}pc + \beta_{3}m_{0}c^{2})^{-2}}{4 E_{p}^{2}} \left[ \frac{2E_{p}m_{0}^{2}c^{2}}{p^{2}}(E_{p}+\alpha_{3}pc + \beta_{3}m_{0}c^{2})^{2}\sigma_{3} \right. \nonumber\\
& \;\;\;\;\;\;\;\;\; \left. + \frac{m_{0}cE_{p}^{2}}{p}(E_{p}+\alpha_{3}pc + \beta_{3}m_{0}c^{2})2i(\alpha_{1}\beta_{2}-\alpha_{2}\beta_{1})\sigma_{0} \right] \nonumber\\
& = \frac{m_{0}^{2}c^{2}}{2p^{2}E_{p}}\sigma_{3} + \frac{m_{0}c}{2p(E_{p}+\alpha_{3}pc + \beta_{3}m_{0}c^{2})}i(\alpha_{1}\beta_{2}-\alpha_{2}\beta_{1})\sigma_{0}
\end{align}
Note that $\{A,B\} = 0$ and $A^{2}=B^{2}=\sigma_{0}$, where $A=\sum_{j=1}^{3}\alpha_{j}\sigma_{j}$ and $B=\sum_{j=1}^{3}\beta_{j}\sigma_{j}$.

\end{document}